\documentclass[prl,aps,epsf]{revtex4}
\usepackage[english]{babel}
\usepackage{graphicx}
\usepackage{epsf}
\usepackage{pstricks}

\begin{document}


\title{Factorization and transverse momentum for two-hadron production
in inclusive $e^+\,e^-$ annihilation}

\author{I.V. Anikin, O.V. Teryaev}
\affiliation{Bogoliubov Laboratory of Theoretical Physics,
             JINR, 141980 Dubna, Russia }

\begin{abstract}

\noindent
We study factorization of processes involving two fragmentation functions in the case
of very
small transverse momenta. We consider two-hadron production in inclusive
$e^+ e^-$ annihilation and demonstrate a new simple
and illustrative method of factorization for such processes including leading order $\alpha_S$ corrections.
\vspace{1pc}

PACS numbers: 12.38.-t, 13.85.Ni; 13.88.+e
\end{abstract}
\maketitle

\section{I. Introduction}

The electron-positron annihilation process is one of the basic hard scattering processes
where the composite structure of hadrons is investigated. Especially, the two hadron production in inclusive $e^+\,e^-$
annihilation (THPIA), {\it i.e.} the process
in which 
two hadrons are detected in the final state, attracts a lot of
attention due to
the possibility for both the experimental and theoretical studies of several fragmentation functions,
in particular the chiral-odd Collins functions \cite{Col93}-\cite{Efr-Col}.
Since available experimental data \cite{Efremov1998, SMC, Belle} correspond to substantially
different values of $Q^2$, taking into account the evolution of
fragmentation (in particular, Collins) functions becomes
an important theoretical task.
In turn, the studies of evolution are intimately correlated to the investigation of QCD factorization.
However, it is well-known that the factorization theorems do not hold for
an 
 arbitrary kinematic
regime. Indeed, if a transverse momentum of the produced hadron is of the same order of magnitude as
the large photon virtuality, $p_T\sim Q$, the corresponding hadronic tensor can be factorized and
expressed in terms of the integrated fragmentation functions (see, for instance,
a detailed analysis of $e^+ e^-$ annihilation in \cite{Collins81}). Even if $p_T$ is much smaller
than the large photon virtuality, but on the other hand, is much bigger than the characteristic
hadronic size, $\Lambda_{QCD}$, factorization of the $e^+ e^-$ annihilation
cross section can still be proven with methods similar to those used for
semi-inclusive DIS or Drell-Yan processes \cite{Ellis1978, Collins84, Ji04}.
However, the regime
for which 
the transverse momentum of produced hadron is of the order of $\Lambda_{QCD}$
faces a conceptual problem how to identify the hard subprocess. Therefore, factorization
becomes vague and requires a special care \cite{Ji04}.

Factorization in THPIA, starting from the seminal paper \cite{Altar},
was
only 
studied in the collinear approximation (see also \cite{deFlor}). Besides,
the procedure of integration over the transverse virtual photon momentum
(see \cite{DBoer}) was restricted to Born approximation.

Since the transverse momentum is of crucial importance for the Collins function, we perform
a detailed analysis of its role in THPIA.
We use ideas of \cite{Rad77, ER81} where
it was   shown 
that
integration over the transverse momentum of
a 
produced $\mu^+\mu^-$ pair or
a 
hadron in the Drell-Yan process provides
an effective propagator of
a highly virtual photon and generates the hard subprocess structure.
This approach was recently generalized \cite{Ter04} to include the case of the
(weighted) transverse momentum average of semi-inclusive deep-inelastic scattering (SIDIS)
with transverse momentum dependent fragmentation functions.

In this paper, we develop these ideas further and present a new method of factorization which may be applied for
any two-current process. In particular, we demonstrate the application of the method in the case of
$e^+ e^-$ annihilation
when 
two hadrons, belonging to different back-to-back jets,
are produced.

To demonstrate the method we are proposing, we consider, at the first stage, the simplest case of
the spin-independent $k_\perp$-integrated fragmentation functions.
We analyze contributions up to leading order $\alpha_S$ corrections to the hard part to obtain the evolutions of
the corresponding fragmentation functions.
Also, we briefly outline a way to extend our approach for the study of the Collins effects.

\section{II. Kinematics}

We consider the process $e^+(l_1)+ e^-(l_2)\to H(P_1)+H(P_2)+ X(P_X)$ where positron and electron carrying
momenta $l_1$ and $l_2$, respectively, annihilate into
a 
time-like photon with momentum $q=l_1+l_2$
for which $q^2=Q^2$ is large.
This time-like photon creates then two outgoing hadrons with
momenta $P_i\, (i=1,\,2)$  (these hadrons belong to two distinct jets) and an undetected bunch
of hadrons with total momentum $P_X$. For such kind of a process, it is convenient to introduce
two invariants
\begin{eqnarray}
\label{zvar}
z_i=\frac{2P_i\cdot q}{Q^2}
\end{eqnarray}
which are analogous to the Bjorken variable.
Note that the variables $z_1$ and $z_2$ are the energy fractions of the detected hadrons in the $e^+\,e^-$
center of mass system. They obey the following constraint due to the energy conservation:
\begin{eqnarray}
\label{cmsrel}
\frac{z_1+z_2}{2} < 1.
\end{eqnarray}
Moreover, due to the momentum conservation there is a stronger constraint on $z_1$ and $z_2$:
\begin{eqnarray}
\label{zreg}
z_1 < 1, \quad z_2 < 1,
\end{eqnarray}
Within this region,  these variables can vary independently.  

To perform the Sudakov decomposition, we choose two dimensionless light-cone basis vectors:
\begin{eqnarray}
\label{lcb}
n^*_{\mu}=(1/\sqrt{2},\,{\bf 0}_T,\,1/\sqrt{2}), \quad n_{\mu}=(1/\sqrt{2},\,{\bf 0}_T,\,-1/\sqrt{2}),
\quad n^*\cdot n =1.
\end{eqnarray}

\noindent
In this paper, we choose the kinematics such that the photon and one of the hadron have purely
longitudinal momenta while the other hadron has both longitudinal and perpendicular momenta:
\begin{eqnarray}
\label{Pframe}
&&P_{2\, \mu}=\frac{z_2 Q}{\sqrt{2}} n^*_{\mu} + \frac{M_2^2}{z_2 Q \sqrt{2}} n_{\mu}, \quad
q_{\mu}=\frac{Q}{\sqrt{2}} n^*_{\mu} + \frac{Q}{\sqrt{2}} n_{\mu},
\nonumber\\
&&P_{1\, \mu}=\frac{z_1 Q}{\sqrt{2}} n_{\mu} +
\frac{M_1^2+\vec{{\bf P}}_{1}^{\perp \,2}}{ z_1 Q \sqrt{2}} n^*_{\mu} +
P_{1\, \mu}^{\perp}.
\end{eqnarray}
The leptonic momenta are
\begin{eqnarray}
\label{lepmom}
&&l_{1\, \mu}=\frac{Q[1-\cos\theta_2]}{2\sqrt{2}}n^*_{\mu} + \frac{Q[1+\cos\theta_2]}{2\sqrt{2}} n_{\mu} +
l_{1\, \mu}^{\perp},
\quad l_{1\, \mu}^{\perp}=\left( \frac{Q}{2}\,\sin\theta_2,\, 0\right);
\nonumber\\
&&l_{2\, \mu}=\frac{Q[1+\cos\theta_2]}{2\sqrt{2}}n^*_{\mu} + \frac{Q[1-\cos\theta_2]}{2\sqrt{2}} n_{\mu} +
l_{2\, \mu}^{\perp},
\quad l_{2\, \mu}^{\perp}=\left( -\frac{Q}{2}\,\sin\theta_2,\, 0\right),
\end{eqnarray}
where $\theta_2$ is the angle between $\vec{P}_2$ and $\vec{l}_1$. This frame is called the $``\perp"$-frame
(or the perpendicular frame) \cite{DBoer, Lev93}. Below, we will omit terms of order $M^2/Q^2$.
Such a kinematics has advantage for analysis of the experimental situation, where the momentum
of one of the produced hadrons is measured.

In order to ensure that the two hadrons are in different jets, we introduce two different variables \cite{Altar}:
\begin{eqnarray}
\label{z2u}
{\cal Z}=\frac{2P_1\cdot q}{Q^2}\equiv z_1, \quad {\cal U}=\frac{P_1\cdot P_2}{P_1\cdot q}.
\end{eqnarray}
However, using eq. (\ref{Pframe}), the difference between ${\cal U}$ and $z_2$ is of the order of $1/Q^2$:
\begin{eqnarray}
{\cal U}=z_2\left[ 1 + \vec{{\bf P}}_{1}^{\perp \,2}/(z^2_1\,Q^2)\right]^{-1}   \  ,
\end{eqnarray}
and can be neglected in the leading order approximation which we are considering in this paper.

The perpendicular projection tensor is defined as usual
\begin{eqnarray}
\label{gP}
g_{\mu\nu}^{\perp}=g_{\mu\nu}-\hat q_\mu \hat q_\nu+\hat T_\mu \hat T_\nu,
\end{eqnarray}
where the two normalized vectors $\hat q$ and $\hat T$ are constructed as
\begin{eqnarray}
\label{norvec}
&&\hat T_{\mu}=\frac{T_{\mu}}{T}, \quad T_{\mu}=P_{2\, \mu}-\frac{P_2\cdot q}{Q^2} q_{\mu},
\nonumber\\
&&\hat q_{\mu}=\frac{q_{\mu}}{Q}.
\end{eqnarray}

\noindent
For simplicity, we consider the unpolarized case.
The differential cross section of the corresponding $e^+ e^-$ annihilation is given by
\begin{eqnarray}
\label{xsec1}
d\sigma(e^+e^-)=\frac{1}{2\,Q^2} \frac{d^3\, \vec{P}_1}{(2\pi)^3\, 2E_1}\,
\frac{d^3\, \vec{P}_2}{(2\pi)^3 \,2E_2}
\sum_X \int \frac{d^3\, \vec{P}_X}{(2\pi)^3 2E_X} \, (2\pi)^4 \,
\delta^{(4)}(q-P_1-P_2-P_X)\, \left| {\cal M} (e^+e^-)\right|^2.
\end{eqnarray}

\noindent
In terms of leptonic and hadronic tensors, we have
\begin{eqnarray}
\label{xsec2}
d\sigma(e^+e^-)=\frac{\alpha^2}{4\,Q^6}\, \frac{d^3\, \vec{P}_1}{E_1}\,\frac{d^3\, \vec{P}_2}{E_2}\,
{\cal L}^{\mu\nu} {\cal W}_{\mu\nu},
\end{eqnarray}
where the hadronic tensor ${\cal W}_{\mu\nu}$ is defined as
\begin{eqnarray}
\label{hadten}
{\cal W}_{\mu\nu}=\sum_X \int \frac{d^3\, \vec{P}_X}{(2\pi)^3 2E_X} \,
\delta^{(4)}(q-P_1-P_2-P_X)\, \langle 0| J_{\mu}(0)|P_1,P_2,P_X\rangle
\langle P_1,P_2,P_X| J_{\nu}(0)|0\rangle.
\end{eqnarray}

We can rewrite the part of the phase space corresponding to the detected hadron with momentum $P_1$ as
\begin{eqnarray}
\label{inphsp}
\frac{d^3\, \vec{P}_1}{(2\pi)^3\, 2E_1}= \frac{d\,z_1}{(2\pi)^3\, 2 z_1} \, d^2\,\vec{{\bf P}}_{1\, \perp}.
\end{eqnarray}
Because the leptonic tensor ${\cal L}_{\mu\nu}$ is independent of the hadronic momentum $P_1$,
it is useful to introduce the averaged hadronic tensor, $\overline{\cal W}_{\mu\nu}^{(\perp)}$,
\begin{eqnarray}
\label{averhadtenPer}
\overline{\cal W}_{\mu\nu}^{(\perp)}&=&\int d^2\, \vec{{\bf P}}_{1\, \perp} \, {\cal W}_{\mu\nu}^{(\perp)}.
\end{eqnarray}

\noindent
Finally,  
 using (\ref{inphsp}) and (\ref{averhadtenPer}),
the cross section (\ref{xsec2}) can be presented as
\begin{eqnarray}
\label{xsecAV}
 E_2\, \frac{d\sigma(e^+e^-)}{d^3\, \vec{P}_2}=
\frac{\alpha^2}{4\, Q^6}\, \frac{d\, z_1}{z_1} \,{\cal L}^{\mu\nu} \overline{{\cal W}}_{\mu\nu}^{(\perp)}.
\end{eqnarray}
Note that this averaging procedure produces the hard subprocess in the Born approximation.

\section{III. Factorization procedure: Born diagram}

Let us now
discuss factorization of hadronic tensors corresponding to
the hadron production in $e^+\, e^-$-annihilation at low $p_T$.
In this case, as pointed out before, a serious conceptual problem is known to be
associated with the difficulty to identify (or, in other words, to separate out)  the hard subprocess.
In \cite{ER81}, it was shown that in the case of Drell-Yan process a suitable integration over
the transverse momentum provides effectively the separation of the hard subprocess.
We extend this approach to the factorization of the
$e^+\, e^-$-annihilation hadronic tensor including leading order $\alpha_S$ corrections.
Furthermore, 
the reproduction of the well-known DGLAP evolution kernel
for both the quark and anti-quark fragmentation functions can be considered as a proof of longitudinal
factorization.

\begin{figure}[htb]
$$\includegraphics[width=12cm]{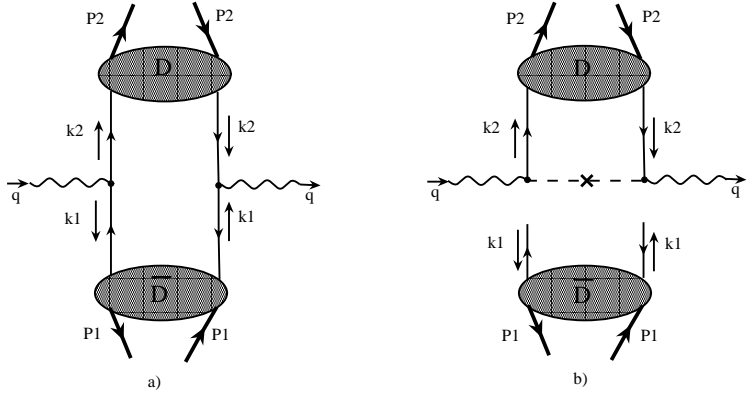}$$
\vspace{-0.5cm}
\caption{ Non-factorized $(a)$ and factorized $(b)$ Born diagrams}
\label{F1}
\end{figure}

First, we consider
a 
simple Born diagram depicted on Fig. \ref{F1} (a).
The corresponding hadronic tensor reads:
\begin{eqnarray}
\label{hadten2}
{\cal W}_{\mu\nu}=\int d^4 k_1\, d^4 k_2\, \delta^{(4)}\left( k_1+k_2-q \right) \,
{\rm tr} \left[\gamma_{\nu} \,\Theta(k_2) \,\gamma_{\mu} \,\bar\Theta(k_1) \right]+ (1\leftrightarrow 2),
\end{eqnarray}
where the four-dimension $\delta$-function, representing the momentum conservation
at the quark-photon vertex, will be treated as the ``hard" part.
The non-perturbative quark and anti-quark
correlation functions $\Theta(k_2)$ and $\bar\Theta(k_1)$ are given
by   
\begin{eqnarray}
\label{FF1}
&&\Theta_{\underline{\alpha\, \beta}}(k_2)=
\int \frac{d^4\, \xi}{(2\pi)^4} \,e^{ik_2\cdot\xi} \,\langle 0| \psi_{\underline{\alpha}}(\xi)|P_2, P_{X_2}\rangle
\langle P_2, P_{X_2}| \bar\psi_{\underline{\beta}}(0)|0 \rangle,
\nonumber\\
&&\bar\Theta_{\underline{\alpha\, \beta}}(k_1)= -
\int \frac{d^4\, \eta}{(2\pi)^4} \,e^{-ik_1\cdot\eta}\, \langle 0| \bar\psi_{\underline{\beta}}(0)|P_1, P_{X_1}\rangle
\langle P_1, P_{X_1}| \psi_{\underline{\alpha}}(\eta)|0 \rangle,
\end{eqnarray}
where the underlined indices are the Dirac indices. The summation and integration over the intermediate
undetected states are implied.
Note that owing to the delta function in (\ref{hadten2}) the hadronic tensor cannot be expressed through
the factorized $k_\perp$-integrated fragmentation functions.
To express the corresponding correlation functions through the fragmentation functions, we
suggest a factorization scheme inspired by the Ellis-Furmanski-Petronzio (EFP) approach \cite{EFP}.
We first consider the formal identical transformation:
\begin{eqnarray}
\label{replace}
&&d^4 k_1 \to d^4 k_1 \, dz^{\prime}_1\,
\delta(P_1^-/k_1^- - z^{\prime}_1)=
d^4 k_1 \, \frac{dz^{\prime}_1}{(z^{\prime}_1)^2}\,
\delta(k_1\cdot \tilde n^*-1/z^{\prime}_1),
\nonumber\\
&&d^4 k_2 \to d^4 k_2 \, dz^{\prime}_2\,
\delta( P_2^+/k_2^+ - z^{\prime}_2)=
d^4 k_2 \, \frac{dz^{\prime}_2}{(z^{\prime}_2)^2}\,
\delta(k_2\cdot \tilde n-1/z^{\prime}_2),
\end{eqnarray}
where the two vectors
\begin{eqnarray}
\tilde n^*_\mu=\frac{n^*_\mu}{P_1\cdot n^*}, \quad  \tilde n_\mu=\frac{n_\mu}{P_2\cdot n}
\end{eqnarray}
have been introduced. Variables $z^{\prime}_i$ may be interpreted as
partonic fractions of
the corresponding momenta of produced hadrons.

As mentioned above,
we treat the four-dimensional $\delta$-function as the hard part of
the  
corresponding tensor.
This assumption will be justified below.
Our
analysis 
is limited by the study of the leading twist contributions.
It means that
we keep only the first terms of the expansion:
\begin{eqnarray}
\label{deltafun}
&&\delta^{(4)}\left( k_1+k_2-q \right)=\delta^{(4)}\left( \frac{P_{1\, \mu}}{z^{\prime}_1}+
\frac{P_{2\, \mu}}{z^{\prime}_2}-q \right)+
O(k_\perp)\simeq
\nonumber\\
&&
\delta\left( \frac{P_2^+}{z^{\prime}_2}- q^+ \right)\, \delta\left( \frac{P_1^-}{z^{\prime}_1}- q^- \right) \,
\delta^{(2)}\left( \frac{\vec{{\bf P}}_1^\perp}{z^{\prime}_1}\right) \  .
\end{eqnarray}
The hadronic tensor is now rewritten as
\begin{eqnarray}
\label{hadtenPer01}
{\cal W}_{\mu\nu}^{(\perp)}=
\int \frac{dz^{\prime}_1}{(z^{\prime}_1)^2}\, \int \frac{dz^{\prime}_2}{(z^{\prime}_2)^2}
\delta\left( \frac{P_2^+}{z^{\prime}_2}- q^+ \right)\, \delta\left( \frac{P_1^-}{z^{\prime}_1}- q^- \right) \,
\delta^{(2)} \left( \frac{\vec{{\bf P}}_1^\perp}{z^{\prime}_1}\right)\,
{\rm tr}\left[
\gamma_\nu \, \Theta(z^{\prime}_2)\, \gamma_\mu \, \bar\Theta(z^{\prime}_1)
\right],
\end{eqnarray}
where
\begin{eqnarray}
\label{Thetaz2}
&&\Theta(z^{\prime}_2)\stackrel{def}{=}
\int d^4 k_2 \,\delta(k_2\cdot \tilde n - 1/z^{\prime}_2) \,\Theta(k_2),
\\
\label{Thetaz1}
&&\bar\Theta(z^{\prime}_1)\stackrel{def}{=}
\int d^4 k_1 \,\delta(k_1\cdot \tilde n^*-1/z^{\prime}_1) \,\bar\Theta(k_1)
\end{eqnarray}

\noindent
Since we study spin-independent fragmentation functions, we first have to project the correlation
functions (\ref{Thetaz2}) and (\ref{Thetaz1}) onto the corresponding
Lorentz vector structures. Starting from the quark
correlator function, we write
\begin{eqnarray}
\label{qcorrfun}
\Theta(z^{\prime}_2)\Longrightarrow  \frac{1}{4}
{\rm tr}\left[ \gamma_\alpha \,\Theta(z^{\prime}_2)\right]\, \gamma_\alpha
\end{eqnarray}
Using (\ref{FF1}) and the integral representation of the $\delta$-function in (\ref{Thetaz2}),
the vector projection of the quark correlation function (see (\ref{qcorrfun})) can be represented in terms of the
spin-independent fragmentation function as
\begin{eqnarray}
\label{D2fun}
&&\frac{1}{4}
\int \frac{d\lambda_2}{2\pi} \,e^{i\lambda_2 /z^{\prime}_2} \,
\int d\, \xi^+ \, d\, \xi^- \,d^2\, \vec{\xi}_\perp\,
\delta(\lambda \tilde n^- - \xi^-) \,
\delta(\xi^+)\, \delta^{(2)}(\vec{\xi}_\perp)
\nonumber\\
 &&{\rm tr} [\gamma_{\alpha} \, \langle 0| \psi(\xi^+, \xi^-, \vec{\xi}_\perp)|P_2, P_{X_2}\rangle
\langle P_2, P_{X_2}| \bar\psi(0)|0 \rangle ] \, \gamma_{\alpha}
=\frac{D(z^{\prime}_2)}{z^{\prime}_2}\, \hat P_2.
\end{eqnarray}
Here, the fragmentation function $D(z^{\prime}_2)$ can also be written as
\begin{eqnarray}
\label{D2fun2}
D(z^{\prime}_2)= \frac{z^{\prime}_2}{4(2\pi)}\int d\xi^- \,e^{i P^+_2 \xi^-/z^{\prime}_2} \,
\, {\rm tr} [\gamma^+ \, \langle 0| \psi(0,\xi^-,\vec{{\bf 0}}_\perp)|P_2, P_{X_2}\rangle
\langle P_2, P_{X_2}| \bar\psi(0)|0 \rangle ]   \ ,
\end{eqnarray}
provided
that 
the minus co-ordinate component $\xi^-$ is equal to $\lambda_2 \tilde n^-$.
Note that this fragmentation function
obeys the momentum conservation sum rules \cite{Collins81,Coll-Sop81, Lev93}:
\begin{eqnarray}
\label{sr}
\sum_{h,\, s} \int dz\,z \,D(z) = 1.
\end{eqnarray}

\noindent
In a similar manner, we project the anti-quark correlation function:
\begin{eqnarray}
\label{D1fun}
&&\frac{1}{4}{\rm tr}\left[ \gamma_\alpha \, \bar\Theta(z^\prime_1)\right] \gamma_\alpha=
\frac{1}{4} \int \frac{d\lambda_1}{2\pi} e^{-i\lambda_1 /z^{\prime}_1}
\int d\eta^+\, d\eta^-\,d^2 \vec{\eta}_\perp \,
\delta(\lambda \tilde n^{*\, +} - \eta^+)\,
\delta(\eta^-)\, \delta^{(2)}(\vec{\eta}_\perp)
\nonumber\\
&&{\rm tr} [\gamma_{\alpha} \,\langle 0| \bar\psi(0)|P_1, P_{X_1}\rangle
\langle P_1, P_{X_1}| \psi(\eta^+,\eta^-,\vec{\eta}_\perp)|0 \rangle ] \, \gamma_{\alpha}
=\frac{\bar D(z^{\prime}_1)}{z^{\prime}_1}\, \hat P_1.
\end{eqnarray}

\noindent
Taking into account
Eqs.  
(\ref{qcorrfun})-(\ref{D1fun}), we bring the hadronic tensor in the following form:
\begin{eqnarray}
\label{hadtenPer2}
{\cal W}_{\mu\nu}^{(\perp)}&=&
\int \frac{dz^{\prime}_1}{(z^{\prime}_1)^3}\, \bar D(z^{\prime}_1)
\int \frac{dz^{\prime}_2}{(z^{\prime}_2)^3} \, D(z^{\prime}_2)
\nonumber\\
&& \delta(P_1^-/z^{\prime}_1 - q^-) \delta (P_2^+/z^{\prime}_2-q^+)
\delta^{(2)}(\vec{{\bf P}}_1^\perp/z^{\prime}_1) \,
{\rm tr} \left[\gamma_{\nu} \,\hat P_2 \,\gamma_{\mu} \,\hat P_1 \right].
\end{eqnarray}
\noindent
Now,   
inserting $P_2^+$ and $P_1^-$ defined via the kinematical variables $z_2$ and $z_1$ (see (\ref{Pframe}))
and calculating the trace in (\ref{hadtenPer2}), we get:
\begin{eqnarray}
\label{hadtenPer3}
{\cal W}_{\mu\nu}^{(\perp)}&=&
-4 g_{\mu\nu}^\perp \, \delta^{(2)}(\vec{{\bf P}}_1^\perp) \,
\biggl[ z_1 \int \,dz^{\prime}_1 \, \bar D(z^{\prime}_1)
\delta(z_1-z^{\prime}_1)\biggr]
\biggl[ z_2 \int \frac{dz^{\prime}_2}{(z^{\prime}_2)^2}\, D(z^{\prime}_2)
\delta(z_2-z^{\prime}_2)\biggr],
\nonumber\\
&=&-4 g_{\mu\nu}^\perp\, \delta^{(2)}(\vec{{\bf P}}_1^\perp) \,
\left[z_1 \bar D(z_1)\right] \, \left[ \frac{D(z_2)}{z_2} \right].
\end{eqnarray}
From (\ref{hadtenPer3}), one can see that, though the hadronic tensor ${\cal W}_{\mu\nu}^{(\perp)}$
has a formally factorized form, it does not have much sense because of the two-dimensional $\delta$-function.
To eliminate it, we have to integrate over the perpendicular momentum or, in other words, to
go over to the averaged hadronic tensor (\ref{averhadtenPer}),
\begin{eqnarray}
\label{averhadtenPer3}
\overline{{\cal W}}_{\mu\nu}^{(\perp)}=
-4 g_{\mu\nu}^\perp \,
\int \,d^2\,  \vec{{\bf P}}_{1\,\perp} \,
\delta^{(2)}(\vec{{\bf P}}_1^\perp) \,
\left[z_1 \bar D(z_1)\right] \, \left[ \frac{D(z_2)}{z_2} \right]=
-4 g_{\mu\nu}^\perp \,
\left[z_1 \bar D(z_1)\right] \, \left[ \frac{D(z_2)}{z_2} \right].
\end{eqnarray}

\noindent
On the other hand, as we will discuss below, integration over $d^2\,  \vec{{\bf P}}_{1\,\perp}$ will
generate the effective diagram depicted in Fig. \ref{F1} (b). Indeed, after integration,
one can represent the one-dimensional $\delta$-function $\delta(z_2-z^{\prime}_2)$ as
the imaginary part of the effective propagator:
\begin{eqnarray}
\delta(z_2-z^{\prime}_2)\Longrightarrow \Im{\rm m}\frac{1}{z_2-z^{\prime}_2}
\Longrightarrow \Im{\rm m}\frac{1}{[q-P_2/z^{\prime}_2]^2}.
\end{eqnarray}
The latter will reduce the diagram in Fig. \ref{F1} (a) to the diagram plotted in Fig. \ref{F1} (b).
Notice that the appearance of this hard effective propagator justifies {\it a posteriori} the
suggested generalization of the EFP factorization scheme.
Namely, the four-dimensional $\delta$-function should be treated as the hard part, and the entering parton momenta
should be replaced, at the leading twist level, by their longitudinal parts.

Let us now discuss the above-mentioned procedure from the viewpoint of the approach of averaging
given by
\begin{eqnarray}
\label{reldd}
d^2\,\vec{{\bf P}}_{1\, \perp}= 2 z_1\, \int\int d^4\, P_1 \, \delta(P_1^2)\,
\delta \left(\frac{2P_1\cdot q}{Q^2}-z_1\right).
\end{eqnarray}
Indeed, integrating over $d^4\, P_1$ with the four-dimensional $\delta$-functions
(see (\ref{hadtenPer2})), one can observe that the one-dimensional $\delta$-function which is
responsible for the mass-shell condition goes to the $\delta$-function
that 
can be understood as the imaginary part of some effective ``propagator" with large
photon virtuality, {\it i.e.}
\begin{eqnarray}
\delta(P_1^2) \Longrightarrow \delta([q-P_2/z^{\prime}_2]^2)\sim
\Im{\rm m}\frac{1}{[q-P_2/z^{\prime}_2]^2}.
\end{eqnarray}
Diagrammatically, it corresponds to the case when the Born diagram transforms to the diagram
plotted in Fig. \ref{F1} (b). The dashed line implies the effective propagator or
the factorization link. So, the averaged hadronic tensor can now be written in terms of the
factorization link,
\begin{eqnarray}
\label{averhadtenPer4}
\overline{\cal W}_{\mu\nu}^{(\perp)}=\int \,d^2\,  \vec{{\bf P}}_{1\,\perp} \,  {\cal W}_{\mu\nu}^{(\perp)}=
\frac{4g_{\mu\nu}^\perp}{\pi} \,\left[ z_1 \bar D(z_1)\right]
\biggl[ z_2 \int \frac{dz^{\prime}_2}{(z^{\prime}_2)^2}\, D(z^{\prime}_2)
\Im{\rm m}\frac{1}{z_2-z^{\prime}_2}\biggr].
\end{eqnarray}

\noindent
The spuriously asymmetric form of (\ref{averhadtenPer3}) or (\ref{averhadtenPer4}) with respect to the
inter changing of $z_1$ and $z_2$ emerges
because of the integration over the momentum $P_1$ of one of the produced hadrons.
Restoring the flavour dependence omitted above and by inserting (\ref{averhadtenPer3}) into (\ref{xsecAV}), we get:
\begin{eqnarray}
\label{xsecAV2}
\frac{d\sigma(e^+e^-)}{d\,z_1\,d\,z_2\, d\cos\theta_2 }=
\frac{3\,\pi\,\alpha^2}{2\, Q^2}\, (1+\cos^2\theta_2)\,
\sum\limits_{a,\bar a} e^2_a \,\bar D^a(z_1) \, D^a(z_2),
\end{eqnarray}
where the only remaining asymmetry is reflected in the polar angle of the detected hadron.
After integration over $\theta_2$, the result becomes completely symmetric:
\begin{eqnarray}
\label{xsecAV3}
\int d\cos\theta_2\, \frac{d\sigma(e^+e^-)}{d\,z_1\,d\,z_2\, d\cos\theta_2}=
\frac{4\pi\,\alpha^2}{Q^4}\, \sum\limits_{a,\bar a} e^2_a \,\bar D^a(z_1) \, D^a(z_2).
\end{eqnarray}

\noindent
Concluding 
this section, we would like to stress that the factorization of the Born diagram
can be implemented in a similar way as in the case of Drell-Yan process \cite{DBoer}. In this case,
we transform the two-dimensional integration $d^2\,\vec{{\bf P}}_{1\, \perp}$ in the
phase space (see (\ref{inphsp})) to the two-dimensional integration over the photon transverse momentum
$d^2\,\vec{{\bf q}}_{T}$ using the well-known relation:
$d^2\,\vec{{\bf P}}_{1\, \perp}=z^2_1 d^2\,\vec{{\bf q}}_{T}$ \cite{DBoer}.
However, the leptonic tensor ${\cal L}^{\mu\nu}$ depends not only on $Q^2$ but also on $\cos\theta_2$.
Therefore, factorization with the transformation to the integration over $d^2\,\vec{{\bf q}}_{T}$ does
not allow for
calculating the angular dependence of the corresponding cross section.

\section{IV. Factorization procedure: leading order $\alpha_S$ corrections}

Now we proceed with the analysis of leading $\alpha_S$ corrections
which are associated with the diagrams that involve gluon emission. It
is known that large logarithms appearing in this case can be
absorbed into the corresponding evolved fragmentation functions.
We will pay a special attention to the terms with mass
singularities which are extracted from the diagrams with the
emission of
real gluons. The corresponding Feynman diagrams
with $\alpha_S$ corrections are depicted in Fig. \ref{F3}. The
domain of integration over the loop momentum in each diagram is in the
region where the considered parton is collinear to either $P_1$
or $P_2$ momentum directions. For the sake of definiteness, we
assume that the hadron with momentum $P_2$ belongs to the quark
jet, while the hadron with the momentum $P_1$ -- to the anti-quark
jet. The opposite situation is trivially obtained by the
interchange of the labels.

\subsection{IVa. Evolution of the quark fragmentation function}

\begin{figure}[htb]
$$\includegraphics[width=12cm]{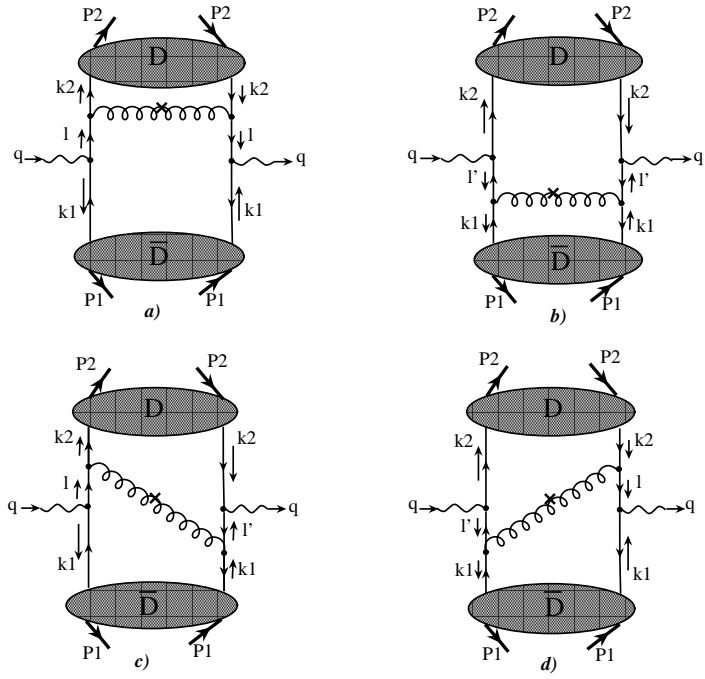}$$
\vspace{-1.cm}
\caption{Leading order $\alpha_S$ diagrams.}
\label{F3}
\end{figure}

To begin with, let us consider evolution of the quark fragmentation function $D(z)$.
At the Born level, the quark which decays into the detected hadron with the momentum $P_2$ and a bunch of
undetected hadrons can emit a real gluon
before and after interaction with the virtual photon.
The diagram depicted in Fig. \ref{F3} (a) corresponds to the interaction of quark with
the virtual photon before the emission of the real gluon.
Choosing Feynman gauge for the gluon fields, we write the hadronic tensor corresponding to this diagram in the form
resembling the Born diagram tensor:
\begin{eqnarray}
\label{NLOa}
{\cal W}_{\mu\nu}^{(\perp),\, q}({\rm Fig.\ref{F3}(a)})&=& g^2 C_F
\int d^4 k_1\, d^4 p\, \delta^{(4)}\left( k_1+p-q \right) \,
{\rm tr} \left[\gamma_{\nu} \,\bar\Theta(k_1) \,\gamma_{\mu} \,
\Omega(p) \right].
\end{eqnarray}
However, instead of the quark correlator $\Theta(k_2)$, we have here the modified correlator $\Omega(p)$
defined as
\begin{eqnarray}
\label{Omega}
\Omega(p)=\frac{\hat p}{p^2} \, \gamma_\alpha
\int \frac{d^4 k_2}{(2\pi)^3}\, \delta([p-k_2]^2)\, \Theta(k_2)
\gamma_\alpha \, \frac{\hat p}{p^2}.
\end{eqnarray}
In principle, factorization of (\ref{Omega}) can be implemented in a similar way.
However, in difference with the Born diagram, the modified tensor (\ref{Omega}) in (\ref{NLOa})
is not completely soft because of the $p$-dependence.

To factorize (\ref{NLOa}), let us first express the parton momenta in terms of the corresponding fractions of
the hadron momenta by means of the integral representation of unity.
The definitions of the fractions for parton momenta $k_1$ and $k_2$ are the same as in (\ref{replace})
whereas for the loop momentum $p$ we write
\begin{eqnarray}
\label{replace2}
&&d^4 p \to
d^4 p \, \frac{d\,y^{\prime}}{(y^{\prime})^2}\,
\delta\left(p\cdot \tilde n-1/y^{\prime}\right).
\end{eqnarray}

\noindent
The pure soft part of (\ref{Omega}) is associated with the quark correlator function $\Theta(k_2)$.
At the same time, the quark propagators and the $\delta$-function
emerging 
from
the imaginary part of
the 
gluon propagator have to be associated with the hard part of (\ref{Omega}).
Therefore, after expanding the $\delta$-function in (\ref{Omega}) about
the momentum $k_2$ around the direction defined by
the 
hadron momentum $P_2$, we get for the correlator:
\begin{eqnarray}
\label{Omega2}
\Omega(p)=\frac{\hat p}{p^2} \, \gamma_\alpha
\int \frac{d z^{\prime}_2}{(z^{\prime}_2)^2}\,
\delta\left(p^2-\frac{2p\cdot P_2}{z^{\prime}_2} \right)
\, \Theta(z^{\prime}_2)
\gamma_\alpha \, \frac{\hat p}{p^2}.
\end{eqnarray}
From this tensor,
which is   the same  
as in the Born diagram, we single out the spin-independent quark fragmentation
function (see eqs. (\ref{qcorrfun}) and (\ref{D2fun})).
Inserting the tensor (\ref{Omega2}) into the hadronic tensor (\ref{NLOa}), we get:
\begin{eqnarray}
\label{NLOa2}
&&{\cal W}_{\mu\nu}^{(\perp),\, q}({\rm Fig.\ref{F3}(a)})= \frac{g^2 C_F}{4(2\pi)^3}
\int \frac{dz^{\prime}_1}{(z^{\prime}_1)^2}\,
\int \frac{d\,y^{\prime}}{(y^{\prime})^2}\,
\delta\left(\frac{P_2^+}{y^{\prime}}-q^+ \right) \delta \left(\frac{P_1^-}{z^{\prime}_1}-q^- \right)
\delta^{(2)} \left(\frac{\vec{{\bf P}}_{1\, \perp}}{z^{\prime}_1} \right) \,
\int \frac{dz^{\prime}_2}{(z^{\prime}_2)^2}
\nonumber\\
&&\int dp^+\, \int d^2\,\vec{{\bf p}}_{\perp}\,
\delta\left( \frac{1}{y^{\prime}}-\frac{p^+}{P_2^+} \right) \, \,
\frac{p^+ - P_2^+/z^{\prime}_2}{p^4_\perp (P_2^+/z^{\prime}_2)^2} \, \,
{\rm tr} \left[ \gamma_{\nu} \, \bar\Theta(z^{\prime}_1)
\,\gamma_{\mu} \, \hat p\, \gamma_\alpha\, \Theta(z^{\prime}_2) \,
\gamma_\alpha\, \hat p \right] \biggl|_{p^- \sim 1/Q},
\end{eqnarray}
where we decomposed again the four-dimensional $\delta$-function around the corresponding hadron directions,
\begin{eqnarray}
\label{4dimdelta}
\delta^{(4)}\left( k_1+p-q \right) \Longrightarrow
\delta\left(\frac{P_2^+}{y^{\prime}}-q^+ \right) \delta \left(\frac{P_1^-}{z^{\prime}_1}-q^- \right)
\delta^{(2)} \left(\frac{\vec{{\bf P}}_{1\, \perp}}{z^{\prime}_1} \right),
\end{eqnarray}
and calculated the integral over $dp^-$ with the $\delta$-function coming from the imaginary part of the gluon
propagator which fixes the minus component of the loop momentum:
\begin{eqnarray}
\label{minp}
\delta([p-k_2]^2)\Longrightarrow
\frac{1}{2[p^+-P_2^+/z^{\prime}_2]}
\delta\left( p^- + \frac{p^2_\perp}{2[p^+-P_2^+/z^{\prime}_2]}\right).
\end{eqnarray}
We remind that $p^+$ and $P^+_2$ are large vectors. Therefore, as one can see from (\ref{minp}),
the minus component is thus suppressed as $1/Q$ and can be discarded in (\ref{NLOa2}).

In (\ref{NLOa2}), we still have
integration over the plus and perpendicular components of
the loop momentum.
Just like in 
Fig. \ref{F3} (a), the quark with momentum $p$ emits a real gluon
and transforms into the quark with momentum $k_2$. This means that
\begin{eqnarray}
\label{loopdec}
p^+= \xi \, k_2^+, \quad dp^+=\frac{P_2^+}{z^{\prime}_2}\, d\xi
\end{eqnarray}

\begin{figure}[htb]
$$\includegraphics[width=12cm]{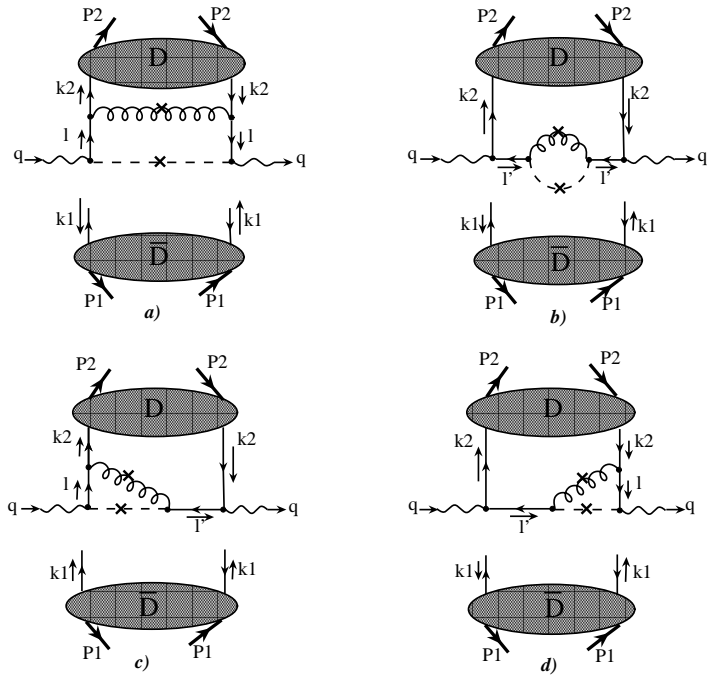}$$
\vspace{-1.cm}
\caption{Factorized leading order $\alpha_S$ diagrams.}
\label{F4}
\end{figure}

\noindent
Using (\ref{loopdec}) and calculating the corresponding integral and the trace
in (\ref{NLOa}), we get the following contribution of the diagram \ref{F3} (a) to the hadronic tensor:
\begin{eqnarray}
\label{NLOaFac}
&&{\cal W}_{\mu\nu}^{(\perp),\, q}({\rm Fig.\ref{F3}(a)})=
-2\, g_{\mu\nu}^\perp\, \frac{\alpha_S}{\pi^2}\, C_F \, \delta^{(2)}\left( \vec{{\bf P}}_{1\, \perp} \right)
\left[ z_1 \bar D(z_1)\right]
\nonumber\\
&&\left[ \int \frac{dz^{\prime}_2}{(z^{\prime}_2)^2} \, D(z^{\prime}_2)\,
\int d\xi \, \delta\left( \xi - \frac{z^{\prime}_2}{z_2}\right)
\left(1-\xi \right) \int \,
\frac{d^2\, \vec{{\bf p}}_\perp}{\vec{{\bf p}}_\perp^2}\right].
\end{eqnarray}
As in the
case of the  
Born diagram, one can show that the $\delta$-function $\delta(\xi-z^\prime_2/z_2)$ in (\ref{NLOaFac})
should be associated with the imaginary part of the hard effective propagator, see
\ref{F4} (a). Note that similar arguments are valid for other diagrams needed for
the study of the quark fragmentation function evolution.

Integration over the two-component loop momentum $\vec{{\bf p}}_T$ should be implemented
with the lower limits defined by the infrared cut-off $\lambda^2$
and the upper limit of $Q^2$. Therefore, the factor
${\rm ln}(Q^2/\lambda^2)$ which appears after this integration reflects
the collinear singularity.

Turning back to the averaged hadronic tensor, we thus derive
\begin{eqnarray}
\label{NLOaFac2}
\overline{\cal W}_{\mu\nu}^{(\perp),\, q}({\rm Fig.\ref{F3}(a)})=
\int \,d^2\,  \vec{{\bf P}}_{1\,\perp} \,  {\cal W}_{\mu\nu}^{(\perp)}=
-2\, g_{\mu\nu}^\perp\, \frac{\alpha_S}{\pi}\, C_F \, {\rm ln}\left(\frac{Q^2}{\lambda^2}\right)\,
\left[ z_1 \bar D(z_1)\right]
\left[ \int \frac{dz^{\prime}_2}{(z^{\prime}_2)^2} \, D(z^{\prime}_2)
\left(1-\frac{z^{\prime}_2}{z_2} \right) \right].
\end{eqnarray}

\noindent
The tensor (\ref{NLOaFac2}) has a completely factorized form and corresponds to \ref{F4} (a) which plays
a role of a ladder diagram.

The diagram \ref{F3} (b) does not contribute to
the terms containing the mass singularity. Indeed, the hadronic tensor corresponding to
such a diagram is given by
\begin{eqnarray}
\label{NLOb}
{\cal W}_{\mu\nu}^{(\perp),\, q}({\rm Fig.\ref{F3}(b)})&=& g^2 C_F
\int d^4 k_1\, d^4 p\, \delta^{(4)}\left( k_1+p-q \right) \,
{\rm tr} \left[\bar\Theta(k_1) \, \Omega_{\mu\nu}(k_1,p) \right],
\end{eqnarray}
where
\begin{eqnarray}
\label{Omegamunu}
\Omega_{\mu\nu}(k_1, p)=\gamma_{\alpha}
\int \frac{d^4 k_2}{(2\pi)^3}\, \delta([p-k_2]^2)\,
\frac{\hat p + \hat k_1 - \hat k_2}{(p + k_1 - k_2)^2}\, \gamma_{\mu}
\Theta(k_2)\, \gamma_{\nu}\, \frac{\hat p + \hat k_1 - \hat k_2}{(p + k_1 - k_2)^2}\,
\gamma_\alpha .
\end{eqnarray}

\noindent
In the same manner as before, one can see that, after integration over $d\,p^-$ with
the $\delta$-function originating from the gluon propagator,
the denominator of (\ref{NLOb}) does not contain the term with $\vec{{\bf p}}_\perp^2$
that produces  
a mass singularity. As a result, this contribution can be discarded.
It is necessary to note that the diagram {\ref{F3}} (b)
effectively corresponds to the diagram Fig. \ref{F4} (b) with the self-energy
insertion into the quark propagator.

Two diagrams presented on \ref{F3} (c) and (d) contribute to the hadronic tensor
in a following way:
\begin{eqnarray}
\label{NLOcd}
{\cal W}_{\mu\nu}^{(\perp),\, q}({\rm Fig.\ref{F3}(c+d)})=2 g^2 C_F
\int d^4 k_1\, d^4 p\, \delta^{(4)}\left( k_1+p-q \right) \,
{\rm tr} \left[ \gamma_{\nu}\, \bar\Theta(k_1) \, \Omega_{\mu}(k_1,p) \right],
\end{eqnarray}
where
\begin{eqnarray}
\label{Omeganu}
\Omega_{\mu}(k_1, p)=\gamma_{\alpha}
\int \frac{d^4 k_2}{(2\pi)^3}\, \delta([p-k_2]^2)\,
\frac{\hat p + \hat k_1 - \hat k_2}{(p + k_1 - k_2)^2}\, \gamma_{\mu}
\Theta(k_2)\, \gamma_{\alpha}\, \frac{\hat p}{p^2}.
\end{eqnarray}
In (\ref{NLOcd}), the denominator contains the necessary power of $\vec{{\bf p}}_\perp^2$ and we have
to keep only the zeroth order
of  
$\vec{{\bf p}}_\perp^2$ in the trace.
Then, following the scheme outlined earlier for other diagrams, we derive
the following expressions for the
hadronic tensor
\begin{eqnarray}
\label{NLOcdFac}
{\cal W}_{\mu\nu}^{(\perp),\, q}({\rm Fig.\ref{F3}(c)+(d)})=
-2\, g_{\mu\nu}^\perp\, \frac{\alpha_S}{\pi^2}\, C_F  \, \delta^{(2)}\left( \vec{{\bf P}}_{1\, \perp} \right)
\left[z_1 \bar D(z_1) \right] \,
\left[\int \frac{dz^{\prime}_2}{(z^{\prime}_2)^2} \, D(z^{\prime}_2)\,
\frac{2\, z^{\prime}_2/z_2}{1-z^{\prime}_2/z_2}\int \,
\frac{d^2\, \vec{{\bf p}}_\perp}{\vec{{\bf p}}_\perp^2} \right];
\end{eqnarray}
and for the averaged hadronic tensor:
\begin{eqnarray}
\label{NLOcdFac2}
\overline{\cal W}_{\mu\nu}^{(\perp),\, q}({\rm Fig.\ref{F3}(c)+(d)})=
-2\, g_{\mu\nu}^\perp\, \frac{\alpha_S}{\pi}\, C_F \, {\rm ln}\left(\frac{Q^2}{\lambda^2}\right) \,
\left[z_1 \bar D(z_1) \right] \,
\left[\int \frac{dz^{\prime}_2}{(z^{\prime}_2)^2} \, D(z^{\prime}_2)\,
\frac{2\, z^{\prime}_2/z_2}{1-z^{\prime}_2/z_2} \right].
\end{eqnarray}

\noindent
Thus, we have all
the 
ingredients for the derivation of the factorized hadronic tensor
including the $\alpha_S$ corrections and mass singularities.
Summing of eqs (\ref{NLOaFac}) and (\ref{NLOcdFac}) together with the contributions related to
the virtual gluon emission gives
\begin{eqnarray}
\label{DEE}
\overline{\cal W}_{\mu\nu}^{(\perp),\, q}=
-2\, g_{\mu\nu}^\perp\, \frac{\alpha_S}{\pi}\, C_F \, {\rm ln}\left(\frac{Q^2}{\lambda^2}\right)
\left[z_1 \bar D(z_1) \right] \,
\left[\int \frac{dz^{\prime}_2}{(z^{\prime}_2)^2} \, D(z^{\prime}_2)\,
\left( \frac{1+(z^{\prime}_2/z_2)^2}{1-z^{\prime}_2/z_2} \right)_+\, \right].
\end{eqnarray}
The factorization scale $\mu_F$ can be introduced by the standard decomposition:
${\rm ln}(Q^2/\lambda^2)={\rm ln}(Q^2/\mu^2_F) + {\rm ln}(\mu_F^2/\lambda^2)$
where the first term should be combined with the hard part of the corresponding hadronic tensor
whereas the second one with the soft part.
If we choose $\mu_F^2=Q^2$, the sum of (\ref{DEE}) with the contribution of the Born diagram
(\ref{averhadtenPer3}) leads to
the following substitution for the quark fragmentation function:
\begin{eqnarray}
\label{DQ2}
D(z_2) \Longrightarrow D(z_2)+ \frac{\alpha_S}{2\pi}\, C_F\, {\rm ln}\left(\frac{Q^2}{\lambda^2}\right) \,
\int\limits_{z_2}^1 \frac{dy_2}{y_2} \, D(z_2/y_2)\, \left(\frac{1+y_2^2}{1-y_2}\right)_+.
\end{eqnarray}
As a result, the quark fragmentation function
acquires $Q^2$-dependence and satisfies the DGLAP evolution equation:
\begin{eqnarray}
\label{DGLAPq}
\frac{d \, D(z_2)}{d\, {\rm ln} \,Q^2}= \int\limits_{z_2}^1 \frac{dy_2}{y_2} \, D(z_2/y_2)\,
V_{qq}(y_2), \quad
V_{qq}(y)=\frac{\alpha_S}{2\pi}\, C_F\, \left(\frac{1+y^2}{1-y}\right)_+ .
\end{eqnarray}

\subsection{IVb. Evolution of the anti-quark fragmentation function}

We will now consider the anti-quark fragmentation functions. The
anti-quark sector in the fragmentation can be studied in
similarly to the quark fragmentation function case. However, there
are some minor differences in comparison to the quark
fragmentation function evolution. First of all, the diagram \ref{F3} (b) will now play a role of the ladder
diagram. Let us write down the hadronic tensor corresponding to
this diagram:
\begin{eqnarray}
\label{antiNLOb}
{\cal W}_{\mu\nu}^{(\perp),\, \bar q}({\rm Fig.\ref{F3}(b)})&=& g^2 C_F
\int d^4 k_2\, d^4 m\, \delta^{(4)}\left( k_2+m-q \right) \,
{\rm tr} \left[\gamma_{\mu} \, \Theta(k_2) \,\gamma_{\nu} \,
\bar\Omega(m) \right],
\end{eqnarray}
where
\begin{eqnarray}
\label{antiOmega}
\bar\Omega(m)=\frac{\hat m}{m^2} \, \gamma_\alpha
\int \frac{d^4 k_1}{(2\pi)^3}\, \delta([m-k_1]^2)\, \bar\Theta(k_1)
\gamma_\alpha \, \frac{\hat m}{m^2}.
\end{eqnarray}

\noindent
As in the case of the quark sector, we first introduce the definition of the corresponding parton fractions
(see  eqs. (\ref{replace}) and (\ref{replace2})). Then, the corresponding $\delta$-functions have to be
decomposed in the directions defined by the hadron momenta. We obtain
\begin{eqnarray}
\label{del1}
\delta^{(4)}\left( k_2+m-q \right) \Longrightarrow
\delta\left(m^+ + \frac{P_2^+}{z^{\prime}_2} -q^+\right) \,
\delta\left(\frac{P_1^-}{y^{\prime}} -q^-\right) \,
\delta^{(2)}\left( \frac{\vec{{\bf P}}_1^{\perp}}{y^{\prime}}\right)
\end{eqnarray}
for the four-dimensional $\delta$-function which is responsible for the momentum conservation at the local vertex:
\begin{eqnarray}
\label{del2}
\delta([m-k_1]^2) \Longrightarrow
\frac{1}{2[m^- - P^-_1/z^{\prime}_1]}
\delta\left( m^+ + \frac{m^2_{\perp}}{2[m^- - P^-_1/z^{\prime}_1]} \right)
\end{eqnarray}
for the $\delta$-function in the tensor (\ref{antiOmega}).
Integration over $dm^+$ with $\delta$-function (\ref{del2}) fixes the plus component
of the loop momentum to be a small variable. As result, $m^+$ can be neglected in the corresponding expressions.

Omitting the details,
since 
all stages of calculations are exactly the same as for the quark sector. Below, we write
the expression for the averaged hadronic tensor corresponding to Fig. \ref{F3} (b):
\begin{eqnarray}
\label{antiqFF4av}
\overline{\cal W}_{\mu\nu}^{(\perp),\, \bar q}({\rm Fig.\ref{F3}(b)})&=&
-2\, g_{\mu\nu}^\perp\, \frac{\alpha_S}{\pi}\, C_F \, {\rm ln}\left(\frac{Q^2}{\lambda^2}\right)
\left[\frac{D(z_2)}{z_2} \right]
\left[z_1^2\int \frac{dz^{\prime}_1}{(z^{\prime}_1)^2}\, \bar D(z^{\prime}_1)
\left( 1-\frac{z_1^{\prime}}{z_1} \right) \right].
\end{eqnarray}

\noindent
The next non-zero contribution comes from the diagrams
presented on Fig. \ref{F3} (c) and (d). These diagrams give us the
following expression:
\begin{eqnarray}
\label{antiNLOcd}
{\cal W}_{\mu\nu}^{(\perp),\, \bar q}({\rm Fig.\ref{F3}(c+d)})&=& 2 g^2 C_F
\int d^4 k_2\, d^4 m\, \delta^{(4)}\left( k_2+m-q \right) \,
{\rm tr} \left[\Theta(k_2) \,\gamma_{\nu} \, \bar\Omega_{\mu}(k_2, m) \right],
\end{eqnarray}
where
\begin{eqnarray}
\label{antiOmegacd}
\bar\Omega_{\mu}(k_2, m)=\frac{\hat m}{m^2} \, \gamma_\alpha
\int \frac{d^4 k_1}{(2\pi)^3}\, \delta([m-k_1]^2)\, \bar\Theta(k_1)
\gamma_{\mu}\, \frac{\hat m - \hat k_1 + \hat k_2}{( m - k_1 + k_2)^2} \,\gamma_\alpha .
\end{eqnarray}

\noindent
Calculating the tensor (\ref{antiNLOcd}) in a way similar to the derivation of (\ref{NLOcd}), we obtain
the averaged hadronic tensor:
\begin{eqnarray}
\label{antiqFFcdav}
\overline{\cal W}_{\mu\nu}^{(\perp),\, \bar q}({\rm Fig.\ref{F3}(c)+(d)})=
-2\, g_{\mu\nu}^\perp\, \frac{\alpha_S}{\pi}\, C_F \, {\rm ln}\left(\frac{Q^2}{\lambda^2}\right)
\left[\frac{D(z_2)}{z_2} \right]
\left[z_1^2\int \frac{dz^{\prime}_1}{(z^{\prime}_1)^2}\, \bar D(z^{\prime}_1)
\frac{2 z_1^{\prime}/z_1}{1-z_1^{\prime}/z_1} \right].
\end{eqnarray}

\noindent
Diagram \ref{F3} (a) in this case does not contribute to the evolution of the anti-quark
fragmentation function due to the same arguments as we presented for the quark sector discussing
the diagram \ref{F3} (b).

Thus, combining all the diagrams and adding the
contributions from the virtual gluon emissions,
we 
get:
\begin{eqnarray}
\label{DbarEE}
\overline{\cal W}_{\mu\nu}^{(\perp),\, \bar q}=
-2\, g_{\mu\nu}^\perp\, \frac{\alpha_S}{\pi}\, C_F \, {\rm ln}\left(\frac{Q^2}{\lambda^2}\right)
\left[z_1^2 \int \frac{dz^{\prime}_1}{(z^{\prime}_1)^2} \, \bar D(z^{\prime}_1)\,
\left(\frac{1+(z^{\prime}_1/z_1)^2}{1-z^{\prime}_1/z_1}\right)_+ \right] \,
\left[\frac{D(z_2)}{z_2} \right].
\end{eqnarray}

\noindent
Again, as for the quark case, the summation of the hadronic tensor (\ref{DbarEE}) with the Born hadronic tensor
modifies the anti-quark fragmentation function:
\begin{eqnarray}
\label{DbarQ2}
\bar D(z_1) \Longrightarrow \bar D(z_1)+ \frac{\alpha_S}{2\pi}\, C_F\, {\rm ln}\left(\frac{Q^2}{\lambda^2}\right) \,
\int\limits_{z_1}^1 \frac{dy_1}{y_1} \, \bar D(z_1/y_1)\, \left(\frac{1+y_1^2}{1-y_1}\right)_+.
\end{eqnarray}
As a result of it, the anti-quark fragmentation function becomes $Q^2$-dependent and obeys the DGLAP evolution
equation (see, (\ref{DGLAPq})).
Besides, the introduction of the hard effective propagators as in the quark case leads to
the factorized Feynman diagrams which are completely analogous to the diagrams on Fig. \ref{F4}.

\section{V. $k_T$-dependent functions: Born approximation}

In the preceding sections, we focused on the case of the
spin-independent integrated fragmentation functions which in a way
was a  test of
our approach. Since we plan to extend our approach to
$k_T$-unintegrated functions (in particular, Collins function) we
discuss the Born approximation of $e^+ e^-$-annihilation, involving both the Collins
fragmentation function and the spin-dependent fragmentation
function. Namely, we consider the production of transverse polarized quark-antiquark pair
associated with $\sigma_{\mu\nu}$-structures in the corresponding matrix elements, see below.

More exactly, the nonperturbative blob (see, for example, Fig. \ref{F1}),
related to the detected
baryon, whose transverse polarization is correlated to the transverse polarization of the quark
can be described by the spin-dependent fragmentation function.
At the same time, in the other blob,
the transition of the transverse polarized antiquark with the
intrinsic transverse momentum into the
unpolarized hadron is described by the Collins fragmentation function.
In latter case, the mentioned transition is related to the azimuthal
asymmetry distribution of hadrons.
Note that the symmetric case, with two similar fragmentation functions, has been well-studied
from both
the theoretical and experimental points of view \cite{Efremov1998, Kumano, Bell, Boer:2008fr}.
The asymmetric situation requires the simultaneous measurement of the meson azimuthal asymmetry
in one jet and baryon (say, $\Lambda$) polarization in another one, Such measurements may be performed
at BELLE \footnote{M. Grosse Perdekamp, private communication.}.

We use the co-ordinate (or the impact  parameter) representation where the explicit
definition of the transverse momentum is not necessary.
In this case, the corresponding hadronic tensor takes the form:
\begin{eqnarray}
\label{HTxspace}
\Delta {\cal W}_{\mu\nu}=\int \frac{d^4 \xi}{(2 \pi)^4} e^{i(q\cdot\xi)}\,
{\rm tr}\left[ \gamma_{\mu} \hat\Theta_1(\xi) \gamma_\nu \hat{\bar\Theta}_2(\xi)\right].
\end{eqnarray}
Here, the spin-dependent fragmentation function has been defined in the co-ordinate space as
\begin{eqnarray}
\label{sp}
\hat\Theta_1(\xi) \Rightarrow \sigma_{\alpha \beta}\gamma_5 P_2^\alpha S^\beta
\int\limits_0^1 \frac{dz_2^\prime}{(z_2^{\prime})^2} e^{-i (P_2\cdot \xi)/{z_2^\prime} }
H_{1\, T}(z_2^\prime),
\end{eqnarray}
where $S$ denotes the hadron transverse polarion,
and the Collins function's analog in the
coordinate space is given \cite{Ter04} by
\begin{eqnarray}
\label{coll}
\hat{\bar\Theta}_2(\xi) \Rightarrow && i M \sigma_{\alpha \beta} P_1^\alpha \xi^\beta
\int\limits_0^1 \frac{dz_1^\prime}{(z_1^{\prime})^2} e^{-i (P_1\cdot \xi)/{z_1^\prime} }
\bar H^{\perp}_1(z_1^\prime),
\quad
\bar H^{\perp}_1(z_1^\prime) = \int d k_T^2 \, \frac {k_T^2 }{M^2}\, \bar H^{\perp}_1(z_1^\prime, k_T^2 ),
\end{eqnarray}
where $\xi$ is the (transverse) position in co-ordinate space; while $M$, introduced due to the dimensional analysis,
implies a parameter of the order of jet mass being the only dimensionful parameter in the soft part.

The position in the coordinate space $\xi$ (see, (\ref{coll}))
yields the derivative over $P_1$ in momentum space. Therefore, inserting (\ref{sp}) and (\ref{coll}) in (\ref{HTxspace}),
we get:
\begin{eqnarray}
\label{HTxspace2}
\Delta {\cal W}_{\mu\nu}=\int\frac{dz_1^\prime}{z_1^{\prime}} \int\frac{dz_2^\prime}{(z_2^{\prime})^2}
\biggl\{ \frac{\partial}{\partial P_1^-} \delta(q^- - \frac{P^-_1}{z_1^{\prime}})\biggr\}
\delta(q^+ - \frac{P^+_2}{z_2^{\prime}}) \delta^{(2)}(\frac{\vec{{\bf P}}_1^\perp}{z_1^{\prime}})
\bar H^{\perp}_1(z_1^\prime) \, H_{1\, T}(z_2^\prime) T_{\mu\nu}(P_1,P_2,S_T),
\end{eqnarray}
where
\begin{eqnarray}
T_{\mu\nu}(P_1,P_2,S_T)={\rm tr}\left[ \gamma_{\mu} (i M \sigma_{+-} P_1^-) \gamma_{\nu}
(\sigma_{-T}\gamma_5 P_2^+ S_T)\right].
\end{eqnarray}
Using the kinematics defined above, we derive
\begin{eqnarray}
\label{HTxspace3}
\Delta {\cal W}_{\mu\nu}= T_{\mu\nu} \delta^{(2)}(\vec{{\bf P}}_1^\perp)
\biggl[ \int dz_1^\prime \left(z_1^\prime \bar H^{\perp}_1(z_1^\prime)\right)
\delta^{(1)}(z_1^\prime -z_1)\biggr]
\biggl[\int\frac{dz_2^\prime}{z_2^{\prime}}H_{1\, T}(z_2^\prime) \delta(z_2^\prime-z_2)\biggr].
\end{eqnarray}
Then, calculating the averaged hadronic tensor, one has the following:
\begin{eqnarray}
\label{averHTxspace}
\overline{\Delta\cal W}_{\mu\nu}= \int \,d^2\,  \vec{{\bf P}}_{1\,\perp} \,
\delta^{(2)}(\vec{{\bf P}}_1^\perp) \Delta {\cal W}_{\mu\nu}=
T_{\mu\nu}
\biggl[ z_1 \bar H^{\perp}_1(z_1) \biggr]^{\prime}
\biggl[\frac{H_{1\, T}(z_2)}{z_2}\biggr].
\end{eqnarray}

Note that the analogous asymmetric combination corresponding to SIDIS
process related to ours by crossing
was
also considered in \cite{Ter04}.
In that case, the incoming quark in SIDIS is described by the transversity
distribution function:
\begin{eqnarray}
\hat h(\eta) =  \sigma_{\mu \nu} \gamma_5 P_1^\mu S^\nu
\int \limits_0^1 dx e^{i x (P_1\cdot\eta) } h (x),
\label{tr}
\end{eqnarray}
where $S_\mu$ is the target polarization. Using (\ref{coll}) and (\ref{tr}), the corresponding
hadronic tensor takes the form:
\begin{eqnarray}
\label{spidis}
\Delta {\cal W}^{\mu \nu} = \int \frac{d^4 \xi}{(2 \pi)^4} e^{-i(q\cdot\xi)} \,
{\rm tr} [\hat h(\xi) \gamma^\mu \hat H(\xi) \gamma^\nu ].
\end{eqnarray}

\noindent
To study the $k_T$ (or $\vec{{\bf P}}_{2\,\perp}$)
distributions in SIDIS and the related asymmetries, we should
consider the weighted hadronic tensor which projects out the
corresponding moment of
the 
Collins function:
\begin{eqnarray}
\label{spidisn}
\Delta_n \overline{{\cal W}}^{\,\mu \nu} = \int d^4 P_2 \delta( P_2^2) (P_2\cdot n_\perp)
\delta \left(\frac{P_1\cdot P_2 }{P_1\cdot q} - z\right) \Delta {\cal W}^{\mu \nu},
\end{eqnarray}
where $n_\perp$ is the unit transverse $4-$vector $(n_\perp\cdot P_1=n_\perp\cdot q=0,\, n_\perp^2=-1)$
which defines the transverse direction.
Using (\ref{coll}), (\ref{tr}) and (\ref{spidisn}), one can see that
the derivative $\partial^\alpha\equiv \partial/\partial P_2^\alpha$ in
\begin{eqnarray}
\label{spiddn}
&&\Delta_n \overline{{\cal W}}^{\,\mu \nu}= i M \int d^4 P_2 \,(P_2\cdot n_\perp)\,
\delta(P_2^2) \,\delta \left(\frac{P_1\cdot P_2}{P_1\cdot q} - z\right)
\nonumber\\
&&\int dx \, dz^{\prime} \partial^\alpha \delta(x P_1+ q -P_2/z^{\prime})
h(x) (z^{\prime} I (z^{\prime}))
{\rm tr} [\gamma_5 \hat P_1 \hat S  \gamma^\mu [\hat P_2 \gamma_\alpha] \gamma^\nu]
\end{eqnarray}
should act only on $(P_2\cdot n_\perp)$, so that
the  result \cite{Ter04} is equal to the standard expression for
the contribution of Collins function, except that the role of the direction of
intrinsic transverse momentum is played by the auxiliary
transverse vector $n_\perp$:
\begin{eqnarray}
{\rm tr} [\hat p_1 \hat S \gamma_5 \gamma^\mu \hat p_3 \hat n_\perp  \gamma^\nu]
\Longrightarrow
{\rm tr} [\hat p_1 \hat S \gamma_5 \gamma^\mu \hat p_3 \hat k_T \gamma^\nu].
\end{eqnarray}
This substitution does not change the azimuthal dependence, as the weighted integration corresponds to azimuthal
average:
\begin{eqnarray}
\langle d \sigma(\phi_h) \cos(\phi_h - \phi_n)\rangle = \cos \phi_n
\langle d \sigma(\phi_h) \cos(\phi_h) \rangle
+ \sin \phi_n \langle d \sigma(\phi_h) \sin(\phi_h)\rangle.
\end{eqnarray}
As a result the azimuthal dependence of the cross-section is transferred to the dependence on the
angle $\phi_n$, and $I(z)$ corresponds to the {\it moment} of the Collins function:
\begin{equation}
I(z)  \sim \int d k_T^2  \frac {k_T^2 }{M^2} H_1(z,k_T^2 ).
\end{equation}

Thus, to describe the Collins effect we suggest the
$k_T$-dependent fragmentation function to be written in the
co-ordinate space, where no specification of intrinsic $k_T$ is
required. Therefore, the calculation of radiation corrections and
evolution may be performed in the same way as in this paper.

Moreover, this approach  may be applied \cite{Ter04} for the
description of higher weighted moments in $k_T$ and $p_T$. Say,
choosing 
the term of order $\xi^2$ in the expansion of unpolarized
fragmentation function in the coordinate space and taking the
$p_T$ moment of the hadronic tensor weighted with $p_T^2$
corresponds to the account of the width of the $k_T$-dependent
unpolarized  fragmentation function. The exponential shape of the
latter, in turn, corresponds to partial resummation \cite{Ter04}
of the infinite series of higher twists, analogous to
that 
leading  to the appearance of non-local vacuum condensates, when
vacuum rather than hadronic averages are considered. The
expansion in coordinate space
accompanied by  
taking the respective weighted
moments of cross-sections provides a complementary definition of
observables, accounting for the shape of $k_T$-dependent
fragmentation (and distribution) functions.

\section{VI. Conclusions}

We described a method which allows us to prove a
factorization of the process with two fragmentation functions.
We would like to point out
that the presented method can be applied for any two-current
process.  
The difficulties of
factorization for such kind of
processes emerge for the case when the kinematical
transversalities inside the hadron are rather small. It leads to
the problem in the definition of what is the hard subprocess for
the process.

Following the idea of the paper \cite{ER81}, we showed that
the corresponding $\delta$-functions in the hadronic tensors
should be treated as the hard parts. It is based on the
observation that these $\delta$-functions can be associated with
the imaginary parts of the effective propagators related to the
well-defined hard subprocess. As a result, we finally have the
completely factorized expression for the hadronic tensor with the
evolved fragmentation functions.

In this paper, the proposed method has been tested in the simplest case when the
differential cross section of $e^+ e^-$ annihilation is related to the spin-independent
integrated fragmentation functions.
We also extended our approach to the study of the spin-dependent
structures and $k_T$-dependent fragmentation functions (Collins function
and transversity fragmentation
function).
This will allow us to perform the leading order QCD fits of relevant experimental data \cite{Kumano}.

\section{Acknowledgments}

\noindent We would like to thank D. Boer, A.V.~Efremov, A.~Bacchetta,
B.~Pire, A.V.~Radyushkin and S.~Taneja for stimulating discussions.
I.V.A. would
like to express the gratitude for very warm hospitality at Ecole
Polytechnique.
This investigation was partially supported by the Heisenberg-Landau
Programme (Grant 2008), the Deutsche Forschungsgemeinschaft under
contract 436RUS113/881/0, the EU-A7 Project \emph{Transversity}, and
the RFBR (Grants 06-02-16215, 08-02-00896,  and 07-02-91557),Russian Federation
Ministry of Science and Education  (Grant MIREA 2.2.2.2.6546), and the EcoNet program.

\vspace{1\baselineskip}

\end{document}